# POWER SAVING STRATEGIES AND TECHNOLOGIES IN NETWORK EQUIPMENT OPPORTUNITIES AND CHALLENGES, RISK AND REWARDS

*Luc Ceuppens, Alan Sardella, Daniel Kharitonov*

Juniper Networks, Inc.

## ABSTRACT

*Drawing from today's best-in-class solutions, we identify power-saving strategies that have succeeded in the past and look forward to new ideas and paradigms. We strongly believe that designing energy-efficient network equipment can be compared to building sports cars – task-oriented, focused and fast. However, unlike track-bound sports cars, ultra-fast and purpose-built silicon yields better energy efficiency when compared to more generic "family sedan" designs that mitigate go-to-market risks by being the masters of many tasks. Thus, we demonstrate that the best opportunities for power savings come via protocol simplification, best-of-breed technology, and silicon and software optimization, to achieve the least amount of processing necessary to move packets. We also look to the future of networking from a new angle, where energy efficiency and environmental concerns are viewed as fundamental design criteria and forces that need to be harnessed to continually create more powerful networking equipment.*

## 1. INTRODUCTION

In a world of ever-improving technology, environmental concerns are starting to gain the recognition they should. And as society embraces eco-responsible manufacturing and fabrication processes, the main focus of green-friendly efforts begin to shift towards energy conservation [1]. Using less energy means using less electricity, which in turn may mean a reduction in greenhouse gas emissions and nuclear waste, depending on how we generate the power. Conversely, energy efficiency increasingly becomes the key metric for designing network equipment. This new metric is driven by a rational desire to decrease operational expenses, protect carriers against rising energy costs, and control the environmental impact associated with energy consumption.

## 2. MEASURING EFFICIENCY

Energy conservation is easy to explain and understand at consumer level. You get better gas mileage in a hybrid car; replacing incandescent bulbs with fluorescent lights has an immediate positive effect on energy bills, and teleworking becomes a green alternative to office-bound workplace [2].

However, it's quite interesting to notice that no one has formalized the energy efficiency criteria for network equipment so far. Various informal parameters are in use today, most popular being the absolute power consumption (in Watts) and a normalized efficiency rating (in Watts/bit).

Unfortunately, none of these ad-hoc metrics are good enough to be used for engineering or research purposes. Absolute power consumption is good for site preparation, but says nothing about efficiency of a router/switch design. Being frugal brings little value if the capacity requirements are not met. On the other hand, basic normalized efficiency rating takes capacity into accommodation, but does not define a unified comparison basis. As routers and switches can be deployed in multiple roles and variable packet touch requirements; it makes little sense to rate Ethernet switches against deep packet inspection devices. Furthermore, even devices with similar processing capabilities can come in different configurations. Redundant routing and forwarding engines and fabric planes may affect the power ratings.

In order to create a formal set of efficiency metrics, we suggest the following approach.

First, we should define a functional area (application) where the metric is going to be used. This can be done for core routing, Ethernet switching, IP/MPLS edge, firewalls, etc.

Second, we should normalize system energy consumption (as function of installed components or units) to the actual (measured) full-duplex throughput.

Therefore, we come to the following formula:

$$E_{CR} = \frac{\sum C(i)}{T}$$

$E_{CR}$ denotes energy consumption normalized to application,
C is the power rating of a router's component,
$i \in I$, where I is the set of all components in configuration,
T is the system's capacity (full-duplex)

Typically, vendors rate the consumption of components (or the entire system) in Watts, assuming the maximum load scenario. Therefore, our energy consumption metric is normally expressed in Watts per Gigabits-per-second.

It is also convenient to normalize this metric to 10*Gbps, which gives a physical reference for the most popular interface type (10GE, as defined in IEEE 802.3ae).

In cases where it is preferable to estimate available packet processing speed at a fixed power budget, the inverse metric can become convenient:

$$E_{ER} = \frac{1}{E_{CR}}$$

Where $E_{ER}$ denotes energy efficiency and is expressed in Gigabits per Watt, or Gigabits per KWatt. For the lack of a better term, we suggest to abbreviate the Gigabit/Kwatt metric to "Gores" (Table 1)

|  | Juniper M40 | M160 | T640 | T1600 | Next-gen |
|---|---|---|---|---|---|
| Slot Capacity, Gpbs | 3.0 | 10 | 40 | 100 |  |
| System Capacity | 40Gbps | 160G | 640G | 1600G |  |
| Technology | 180nm | 180nm | 130nm | 90nm | 65nm and < |
| Max System Draw | 1.5 KW | 3.15 KW | 4.52 KW | 8.35 KW |  |
| EER (Gbps/KW) | 13 Gores | 25 Gores | 71 Gores | 96 Gores | > 100 Gores |
| FRS | 1998 | 2000 | 2002 | 2007 | 2010+ |

Table 1 Sample energy efficiency metric (EER) as a function of router generations with respect to the timeline

The use of $E_{CR}$ or $E_{ER}$ to communicate router/switch energy metrics properly expresses the fact that modern routing systems tend to consume more electricity per rack; in return, they process packets with speeds that were not previously possible. Thus, the goal of building the more efficient platforms can be defined as $E_{CR}$ or $E_{ER}$ optimization.

## 3. WHERE DO WE STAND TODAY?

Overview of the building blocks of current routing systems provide a good starting point to explore a power optimization strategy. Assessing the properties of current and former designs is crucial to the quest for power efficiency.

### 3.1. The Forwarding Path

Design of the packet forwarding path in current-generation routers has come a long way since the use of general-purpose computers that saw service in the early days of the Internet. Evolving from off-the-shelf CPUs to limited hardware assistance and ultimately to silicon-only forwarding, network routers have been able to provide the speed and features required for robust and scalable Internet services.

All things being equal, energy efficiency in the packet forwarding path largely depends on the number of gates in the forwarding hardware. In general, every cell, block and gate—whether used or not—requires power, making a strong case for structural optimization within the forwarding engine.

General-purpose CPUs typically present the worst case with respect to power efficiency. The latest multi-core designs are manufactured with 45 to 65nm technology and can feature over two billion transistors. They are fully programmable and can perform any packet lookup operation in existence, but this comes at a cost of relatively high power consumption. Practically speaking, modern general-purposes CPUs form a good basis for deep-packet processing devices. When coupled with the right software, they can achieve forwarding rates on the order of several gigabits per second within a power budget of 70-90 Watts for a high-end CPU, which translates into 150-300 Watts of continuous power draw per system [3]. This includes the supporting chipset, memory channels and glue logic required to form a functional networking platform.

Despite enhancements like dynamic power management, the extensive overhead associated with blocks, gates and instructions not used for packet processing means that general-purpose CPU-based platforms would have $E_{CR}$ on the order of 400-800 Watts per 10 Gbps. This is adequate for medium-to-high touch packet operations, but can be unreasonably high for simpler tasks like IPv4 routing, service differentiation, and firewall filtering. Using general-purpose CPUs in this context is similar to running a family sedan on the racetrack – it can be done, but progressively makes less and less sense as the speed continues to rise. A family sedan running close to its limits is not only unstable, but can be beaten easily by sports cars in terms of both speed and mileage per gallon.

On the other end of the spectrum, one can choose a very complex custom-built and hand-optimized piece of silicon for packet forwarding. The research and manufacturing costs associated with custom packet engines tend to be high because they have a fairly irregular internal structure and are practically "hand-built" to just move packets within certain feature and speed goals. However, the high development cost can be ultimately offset with superior scaling and higher energy efficiency. The current record for energy efficiency in IP/MPLS routing is held by Juniper T1600, with $E_{CR}$ of 94 Watts/10Gbps in a fully loaded configuration. This is almost an order of magnitude better than CPU-based platforms.

In between these two extremes (off-the shelf CPU and fully custom silicon) there are many intermediate solutions featuring a broad array of price/performance ratios and ranging from packet-optimized network processors (NPUs) to fully configurable CPU arrays [4], where features and instructions can be added or removed at will.

Merchant silicon, in general, tends to be cheaper, which explains why many network vendors gravitate towards using NPUs or configurable CPUs. We estimate the $E_{CR}$ range for NPU-based router designs to scale to 200 to 400 Watts per 10Gbps, while the best examples of configurable CPU array systems can demonstrate $E_{CR}$ on the order of 150 to 200 Watts per 10 Gbps. The drop in power efficiency compared to custom silicon is directly correlated to the rise in the powered gate count. Less targeted and more generic designs require more building blocks to achieve the same goal.

### 3.2. Memory Technologies

Choosing a memory system is one of the biggest technology challenges when designing a modern router. Progress in memory channel bandwidth has trailed Moore's law throughout the last decade, while Internet bandwidth has grown much faster than Moore's law. This has created a significant discrepancy between the demand for memory throughput and the technology that is actually available. Thus, hardware engineers face hard choices between cost, speed, energy efficiency, and the general availability of prospective memory technologies.

The tradeoff for hardware engineers is whether to heavily optimize memory access algorithms to retain flexibility and programmability, or to remove certain features like flexibility and power efficiency in favor of more complex (and limited) linear (i.e., RLDRAM) or non-linear memory structures, such as TCAMs. Unlike trie-based memory lookup tables, TCAMs feature deterministic prefix matching speeds and thus can significantly relax processing requirements in the forwarding path. In fact, some contemporary router designs use TCAM for the execution of both forwarding and filtering, thus effectively freeing packet processors to perform only service differentiation and generic lookup tasks.

However, TCAM advantages do not come for free.

Drawbacks are numerous and include limited capacity, huge power consumption, a fixed order of feature execution and issues with lookup masks crossing bit boundaries [5]. Still, many vendors find this acceptable because TCAM technology allows for lower-grade packet processing designs without complex memory lookup algorithms.

The technology barrier in the lookup area is high enough to prevent most companies to cross it. But while both linear and content-driven lookup approaches are proven in the context of large lookup structures and are widely used, prefix trie-based lookup is generally more efficient with respect to power, which makes it worth the research.

### 3.3. Chipset Integration and Fabrication Process

Fundamentally, the power consumption of an ASIC is related to losses during the transfer of electric charges, which in turn is driven by imperfect conductors and electric insulators. The continuing progress of technology results in significant reductions in the three major sources of transistor leakage: sub-threshold leakage, junction leakage, and gate oxide leakage. The immediate benefits of reduced transistor leakage are lower power and decreased power consumption. In other words, it takes less energy to execute the same number of state transitions as a higher-powered chip would use.

Building on better conductor and insulation technology, shrinking silicon fabrication processes allow for higher performance and density with lower dynamic power consumption and cost. Circuits designed at 90nm will typically consume less power than those at 120 nm.

Another dimension to better fabrication process is the ability to integrate more functions into one chip of the same size. Highly integrated chipsets drastically reduce power consumption as I/O buffers connecting multiple chips can be completely or partially eliminated.

For example, in September 2000, Juniper introduced a compact 5 Gpbs (full-duplex) router commercially known as M10. It was based on a chipset featuring three separate chip types: an IP Processor, an I/O processor, and crossbar fabric. These were all manufactured with IBM SA12/SA12e 0.25μm technology. The power budget for the fully loaded M10 system was 494W. In November 2003, this product was replaced by a newer M10i router, which used the same chipset to power the same bandwidth. However, advances in the fabrication process (SA27e 0.18μm) allowed packaging of all chip types (IP, I/O and crossbar) into a single ASIC. This had an immediate positive impact on the power budget of the system, with improvements extending to better PCB utilization, fewer glue logic connections, and simpler signal propagation and memory subsystems. As a result, M10i could power the configuration identical to that of M10 in an envelope of merely 350.4W, a thirty percent improvement in power.

### 3.4. Packaging and Technology Transition Points

While the idea of using the latest fabrication technology and denser packaging may seem trivial from the researcher's standpoint, it may present significant challenges from the product marketing perspective.

First and foremost, given the length of the design cycles, commercial projects always face a problem of dealing with risks. Since a technology choice is done very early, the decision may bring a lot of risk to the schedule and final product. But while the early technology risks have always been well known and understood, power efficiency brings a whole new dimension to the problem of choosing the right technology and packaging. Traditionally, the ISP CAPEX model assumed technology depreciation over the course of five to seven years. Furthermore, some vendors argued that in the presence of multichassis (MC) router designs, the technology lifecycle should be extended to ten years in production or more, while growing the node capacity with adding the linecard routers (LC) to existing routing cluster.

This presents an interesting dilemma. Although the routing/switching hardware progresses slower than the Moore's law, technology generations still change every three to four years. Furthermore, every five years there appears to be a major structural optimization or architecture, which renders the older designs obsolete. This means that every five years, the energy efficiency metrics can be improved dramatically – provided that customers will trade their existing hardware for newer design (Fig 1)

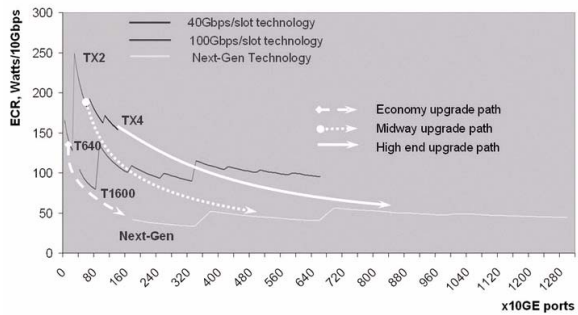

Fig 1 Energy metrics as a function of technology generations

It can be argued that the new hardware can be easily made compatible with the older gear – for example, old and new-generation routers can work together in the same cluster. However, this compatibility comes at a steep price – legacy data structures and architectures have to be maintained, which dramatically limits the potential of the newer technology.

In a sense, energy efficiency forces customers and vendors to accelerated technology introduction, which is significantly different from the situation we saw only five years ago. Today, the cost of ownership for the legacy equipment starts driving the faster network upgrade cycles.

## 4. LESSONS AND CHALLENGES

Looking deeper into the power consumption of alternate router designs, it is easy to notice one trend: simpler and faster packet forwarding silicon achieves the best energy cost per gigabit. In a sense, this is the opposite of the automotive design process — while we are all used to the idea that the fastest sports cars are also the ultimate petrol guzzlers, in the routing world, highly integrated, ultra-fast routers yield the best energy efficiency metrics. Interpolating this trend, one can legitimately argue that the best power efficiency can be achieved by dramatic reduction in packet processing depth with an equivalent increase in the speed and density of the routing platforms.

However, this approach also creates several important challenges. One is the obvious risk of oversimplification. If a router is super-fast but lacks the required functionality and features needed to build a robust and secure communication infrastructure, it will find limited use.

Another issue is related to hardware specialization. Purpose-built silicon yields the best power efficiency, yet can lead to limitations in feature sets. This in turn limits the number of applications the router can serve. Returning to the automotive analogy, purpose-built sports cars do not have the flexibility of a utility truck or a minivan. This creates a risk whenever a need for new functionality arises. To mitigate this risk, some vendors are taking a path of doing relatively complex designs, with a point that the more complex silicon allows for a wider feature set and can ultimately lead to power savings from collapsing multiple network elements into one box [6]. While this approach is certainly promising, it can be also argued that the same task could be carried more efficiently with different types of purpose-built components residing within a single system, allowing for a better mix-and-match of features and power budget to the actual network requirements.

## 5. THE FUTURE OF NETWORKING: OPPORTUNITIES AND RISKS

Looking into the future, we can identify several fascinating trends.

First, it is encouraging to note that increasing power efficiency requirements do not hamper the development of a faster Internet. Indeed the opposite is true – the new set of efficiency metrics supports best-of-breed designs and helps higher speed (next in line: 100 Gbps) packet processor designs to become more widely accepted. Custom silicon designs offer the best tradeoff between features and the resources to run them, and we expect this approach will continue to yield great effectiveness.

Second, the need for power efficiency stimulates fresh thinking in the network data plane area. The progress of networking in the last 25 years has left a great deal of overhead in the form of features, protocols and capabilities that are rarely used and can be dropped as redundant or obsolete. This "de-featuring" trend, combined with topological simplification, could well signal a return to elegance and efficiency in the world of telecommunications.

Finally, while continuing to invest in better memory, fabrication and custom silicon design technologies, we also suggest an eco-friendly approach to generic network design. If a network can be blueprinted to avoid multiplying technologies beyond their actual needs, it can dramatically reduce power consumption by cutting down on large-scale deep packet processing and using denser and faster silicon for routing and switching.

And, after all, even the automotive industry proves that with newest nameplates like Tesla Roadster and Fisker Karma, energy efficiency does not have to be boring. With the right technology and inspiration, we can certainly build faster, simpler and more usable network equipment with smaller environmental footprint.